\documentclass[twocolumn,prl]{revtex4}

\usepackage{graphicx}

\renewcommand{\v}[1]{\ensuremath{\mathbf{#1}}} 

\begin{document}

\title{Water drives peptide conformational transitions}

\author{Dmitry Nerukh}
\email{D.Nerukh@aston.ac.uk}
\affiliation{Non-linearity and Complexity Research Group, Aston University, Birmingham, B4 7ET, UK}

\author{Sergey Karabasov}
\email{sak36@cam.ac.uk}
\affiliation{Whittle Laboratory Cambridge University Engineering Department, 1 JJ Thompson Avenue, Cambridge, CB3 0DY, UK}

\begin{abstract}

Transitions between metastable conformations of a dipeptide are 
investigated using classical molecular dynamics simulation with explicit 
water molecules.  The distribution of the surrounding water at different 
moments before the transitions and the dynamical correlations of water with the 
peptide's configurational motions indicate that water is the main driving 
force of the conformational changes.

\end{abstract}

\maketitle

Investigations of protein dynamics have recently led to believe that water 
plays the major role in protein motion. 
There is a large body of experimental and simulation evidences 
\cite{Born2009, Pagnotta2010, Johnson2010, Zhang2009} showing a close 
connection between the water dynamics and the protein conformations.
Frauenfelder and colleagues have 
experimentally shown that protein dominant conformational motions are 
slaved by the hydration shell and the bulk solvent 
\cite{Frauenfelder2009}, while the protein  
molecule itself provides an `active matrix' necessary for guiding the 
water's dynamics towards biologically relevant conformational changes. 
The change in water dynamics at the  
shell of up to almost a dozen water molecule diameters
around proteins is found in \cite{Ebbinghaus2007}. 

Despite extensive research on protein dynamics 
the investigations of \emph{elementary conformational 
motions} are rare.  Specific molecular mechanisms, 
including the involvement of water molecules, that drive the  
conformational moves are highly demanded as they ultimately define all 
rearrangements of proteins as a whole.

In this work we analyse molecular dynamics (MD) 
simulated peptide focusing on the moments of elementary conformational 
changes including explicit water molecules.  We 
show that water indeed drives the changes, we elucidate the specific 
mechanisms of this phenomenon.

\begin{figure}
\includegraphics[width=\linewidth]{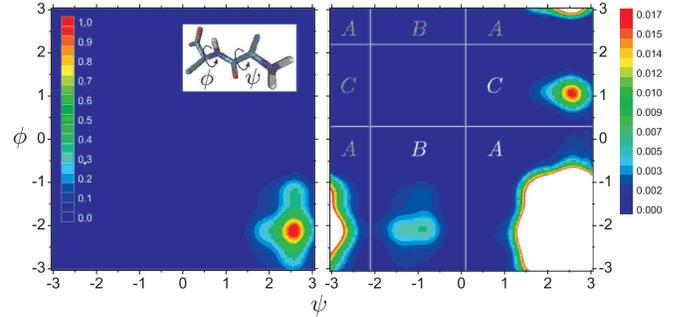}
\caption{\label{fig:psi-phi_histogram}
Left: L-alanyl-L-alanine zwitterion and the normalised density of its 
conformations (Ramachandran plot) formed by a 1$\mu$s trajectory; right:
the same probabilities emphasising the presence of two minor conformations 
and the partitioning for symbolisation}
\end{figure}

We study a zwitterion L-alanyl-L-alanine, Fig.~\ref{fig:psi-phi_histogram},
a very convenient model because i) the conformation of the molecule is
completely defined by the two dihedral angles $\psi$ and $\phi$, ii) in
water the conformation $\psi \approx 2.5$, $\phi \approx -2.2$ radians is
prevalent, however very rare transitions to two other metastable
conformations take place, and iii) the transitions only happen in water
because of the molecule's charged ends.

The three well separated metastable states, clearly visible on the density
of states, Fig.~\ref{fig:psi-phi_histogram}, allow to introduce a simple
natural  discretisation of the conformational states.  By also 
discretising time with a 
step $\Delta t$ the continuous MD trajectory can be converted into a
string of symbols $\{s_i\}, i=0\dots N$, where $s_i$ equals to `$A$',
`$B$', or `$C$' depending on where the trajectory point falls at the time
moment $t_i$, $N$ is the number of time steps.

An important problem is how to identify the moments of transition.  The
described discretisation of the conformational states defines boundaries of
the states.  However, they delineate the density of conformations averaged
over the  whole trajectory.  As we are interested in the dynamically
metastable states, that is the configurations in which the trajectory
spends significantly more time compared to the time it spends in
transitions between the configurations, a \emph{dynamical} model capable of
identifying the transitions is required.  Also, the trajectory does not go
directly from one state to another, instead it winds in a complicated
manner, often crossing the states boundaries many times.


One of the most popular description of protein dynamics lately is the
Markov State Model (MSM) \cite{Schuette1999}.  The model specifies the
probabilities of each  of the discrete states as well as the probabilities
of the transitions between them.  The MSM transition matrix can be
calculated from the MD trajectory by counting the state changes and for the
studied molecule it is
\[
\begin{array}{cccc}
  & A & B & C \\
A & 0.997 & 0.002 & 0.001 \\
B & 0.261 & 0.737 & 0.002 \\
C & 0.097 & 0.001 & 0.901
\end{array}
\]
where the value at row $i$ and column $j$ gives the probability of going to
state $j$ being currently at state $i$.  MSM provides many useful
quantities describing the system  \cite{Schuette1999}. In particular, it
tells that the transition $A \rightarrow B$ happens once in 2.8$ns$ on
average.  We will concentrate on this specific transition for the rest of
the paper.

The moments of the $A \rightarrow B$ transitions within the MSM framework
are the values of $t_i$ for which $s_i$ is equal to $B$ while being equal
to $A$ at the previous time moment $t_{i-1}$.  The time precision of
identifying the transitions is $\Delta t$, which is insufficient for
studying the transitions themselves. The model is valid only for relatively
large time steps, that follows from the requirement for  the transitions to
be history independent (statistically uncorrelated).  For this peptide the
minimal valid time step is $\approx6ps$.  This value is of the same order
as the period of fluctuations within each conformational state and, most
importantly, this is approximately the duration of the process of the
trajectory passing from state to state. Therefore, the MSM has to be
augmented in order to be able to describe the dynamics at significantly
shorter time steps.

For this purpose we build a variant of the \emph{hidden Markov model} using
the same configurational states of the peptide. Specifically, we use the
`$\epsilon$-machine' by Crutchfield et al \cite{Crutchfield1989,
Nerukh2008PRE}.  Instead of the conformational  states, $s_i$, themselves
($A$,$B$,$C$) we  consider the $l$-long \emph{sequences} of states
$\overleftarrow{s}_i \equiv \{s_{i-l+1} \dots s_{i-2} s_{i-1} s_i\}$. The
advantage of such description is that for a small time step, even if the
original states are correlated over several steps, for long enough
sequences $\overleftarrow{s}_i$, these new states (the sequences) are
uncorrelated.  We, therefore, can build a Markov model on these new states.

\begin{figure}
\includegraphics[width=\linewidth]{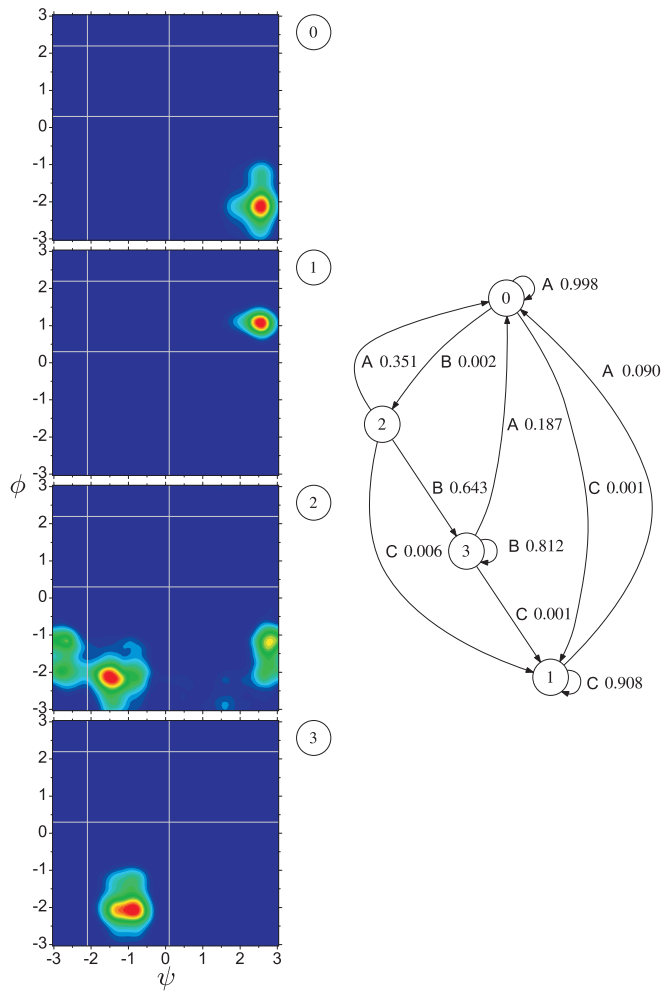}
\caption{\label{fig:5ps_states}
Hidden Markov states for the time step of $5ps$; left: conformations 
corresponding to all time frames belonging to the hidden Markov states; 
right: the hidden Markov states and transitions between them, the labels 
on the arcs indicate the symbol following the state as the result of 
the transition and the transition probability}
\end{figure}

The Markov property on the original states is fulfilled for $\Delta t >
6ps$. For $\Delta t =5ps$ the sequences of at least two time steps are
required to build the hidden Markov model.  The model is shown in
Fig.~\ref{fig:5ps_states}. Here the states `0', `2', and `1' correspond to
the conformational states $A$, $B$, and $C$ since they mostly consist of
the sequences $0\equiv AA$, $2\equiv BB$, and $1\equiv CC$ respectively.
An additional state `2' describes the transition process from $A$ to $B$.
It consists of the sequence $AB$ and has two main transitions from it.
With the probability 0.643 the following symbol is $B$ and the next state
is $3$ which means that the system is transferred to the conformational
state $B$.  There is, however, a significant probability of 0.351 for the
next symbol to be $A$, which describes the return to state 0 or the
original conformation $A$.

This analysis illustrates the advantage of the hidden Markov model: it 
elucidates the \emph{mechanism} of the $A \rightarrow B$ transition.  It 
also explains where the non-Markov property comes from and gives the time 
scale limit at which different pathways of the conformational transition 
start to differentiate from each other.

\begin{figure}
\includegraphics[width=0.65\linewidth]{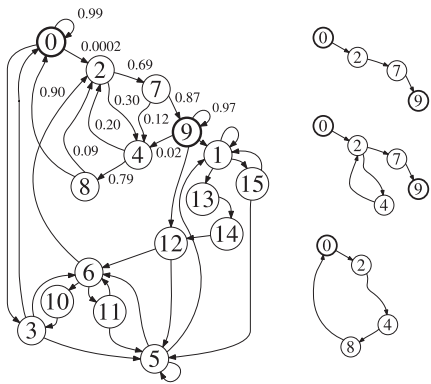}
\caption{\label{fig:e-machine}
Left: the $\epsilon$-machine for the time step $0.3$ ps, the length of the
sequences is $l=4$, state `$0$' corresponds to the original state `$A$'
(mostly consists of the sequences $AAAA$), similarly, state `$9$'
corresponds to `$B$', state `$9$' can only be reached from state `$0$' via
the states  `$2$', `$7$', `$4$', `$8$' that describe the mechanism
(pathways) of the $A \rightarrow B$ transition; right: three  typical cases
during the transition: direct transition (top, probability 0.61),
transition with several recrossings (middle, probability 0.04), failed
attempt of transition (bottom, probability 0.07), the probabilities of
these cases are given assuming that the probability of going from state `0'
to state `9' by any possible route is 1}
\end{figure}

\begin{figure}
\includegraphics[width=0.9\linewidth]{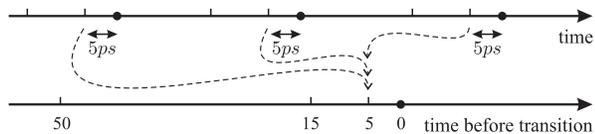}
\caption{\label{fig:timeline}
Collecting the time frames for the `time before transition' statistics; the 
dots on the `time' axes are the transition moments}
\end{figure}

Using the hidden Markov model the time step can be reduced to $0.3ps$, the
model is described in Fig.~\ref{fig:e-machine}. The transitions are the
moments when the system enters state 9 being before in any other state. The
model predicts that the transitions last on average for $\approx 1ps$,
Fig.~\ref{fig:e-machine}, right.  Thus, the $0.3ps$ precision in
identifying them is satisfactory. This gives us a tool to investigate what
happens at different moments \emph{before} the transition.  In particular,
we can study the behaviour of water.  For this we collect the time frames
at specific times before the transitions, Fig.~\ref{fig:timeline}.

We calculate the density of oxygen (hydrogen) atoms \emph{by averaging over
the selected time frames}.  The obtained field $\rho(\v{x},t)$ gives the 
probability of finding 
water atoms at various locations $\v{x}$ around the peptide at 
times $t$ in 
advance of the transition.

\begin{figure}
\includegraphics[width=0.5\linewidth]{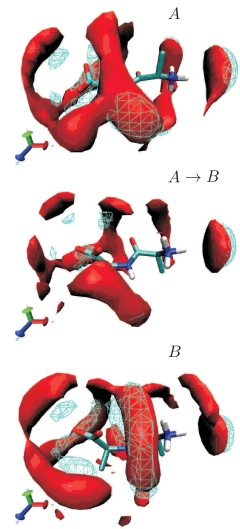}
\caption{
\label{fig:AA_water_3states}
Density isosurfaces at 0.7 $\frac{am}{A^3}$ (average: 0.5) of oxygen (red) 
and 0.1 $\frac{am}{A^3}$ (average: 0.08) of hydrogen (blue) for states $A$ 
(top), $B$ (bottom), and  $\approx 3ps$ before the transition}
\end{figure}

The transition from conformation $A$ to conformation $B$ corresponds to
$\approx180^\circ$ flip of the $NH_3$ group (the right hand side of the
molecule in Fig.~\ref{fig:AA_water_3states}).  Both ends of the molecule
posses charges, negative on the $CO_2$ and positive on the $NH_3$ sites,
which leads to relatively strong attachment of water molecules at the ends.
Hydrogen bonded water molecules to the oxygens and to the hydrogens form
the dense areas of water corresponding to more rigid hydrogen bonds
network, that is more stable structures.

\begin{figure}
\includegraphics[width=0.9\linewidth]{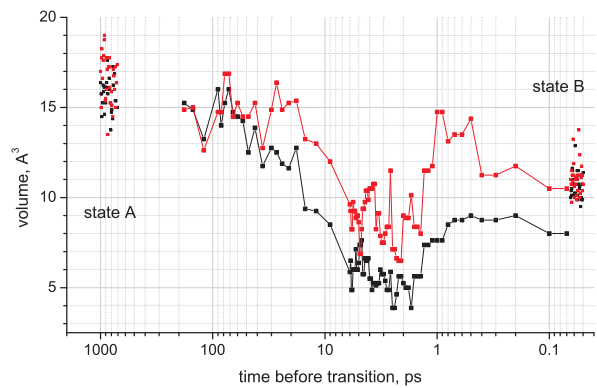}
\caption{
\label{fig:av_density} 
The size of the high density areas (the isosurface values are the same as
in Fig. \ref{fig:AA_water_3states} for oxygen (red) and hydrogen (black))
during the transition from state $A$ to state $B$ (the points for state B
are artificially shifted to the value of $\approx 0.05ps$}
\end{figure}

This is an intuitively clear result.  More interesting is the moment just
before the transition, Fig.~\ref{fig:AA_water_3states}, middle.  Even
though the overall structure of the dense areas remains similar to the $A$
state, evidently, the total area of the dense water is significantly
reduced.  We quantify this decrease by measuring the volume of the dense
areas depicted in Fig.~\ref{fig:AA_water_3states}.  The result as the
function of time before the transition is given in
Fig.~\ref{fig:av_density}.  During the time period when the rotation of the
$NH_3$ group is most significant, from $\approx 9ps$ to $\approx 1ps$, the
size of the dense areas of water is more than 2 times for oxygen and 3
times for hydrogen smaller compared to those during the stable periods of
conformation $A$.  This corresponds to more diffuse character of the
hydrogen bonded network of water molecules as they tend to appear at
different, less concerted locations for different transitions.

However, the most surprising effect following from this analysis is that
the density of water starts reducing as early as $\approx 50ps$ in advance
of the transition.  This is almost 10 times earlier than the actual
conformational change of the peptide!

The above analysis reveals the changes in water densities but it does not
show which part of it is directly correlated with the changes in the
peptide angles. We analyse the dynamical correlations between the dihedral
angles and the water density using the Linear Stochastic
Estimation (LSE) model (originally developed for visualising coherent
structures in turbulent flows \cite{Adrian1994} and the identification of
noise sources in turbulent jets \cite{Kerherve2010}).

\begin{figure}
\includegraphics[width=\linewidth]{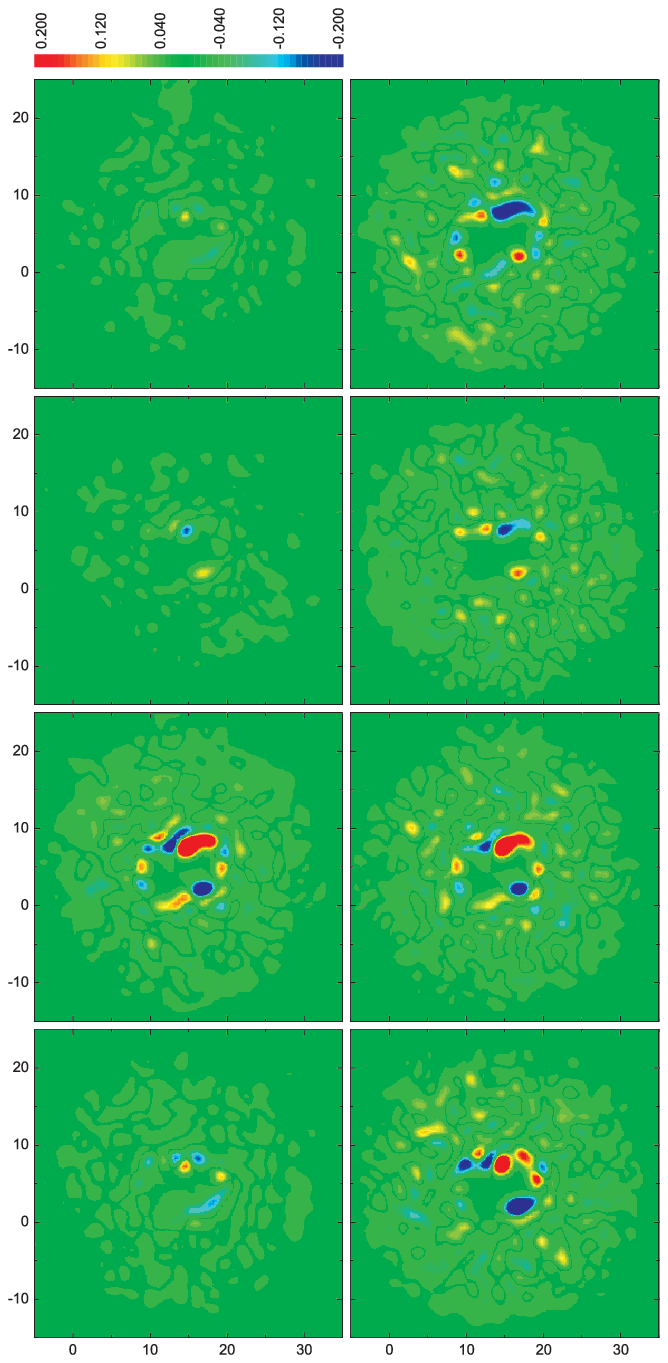}
\caption{
\label{fig:dens_corr} 
The $xy$ cross-section (the value of the $z$ coordinate is chosen such that
the cutting plane passes through the centre of mass of the peptide) of the
density fluctuation field $\rho'(\v{x},t)$ (right), and its part correlated
with the peptide's conformation $\hat{\rho}(\v{x},t)$ (left) for oxygen;
the time before transition are, from top to bottom, $94.8ps$, $21ps$, 
$2.4ps$, and $0ps$; 
the density scale is restricted to the $-0.2\dots 0.2$ interval for clarity,
the maxima of the peaks reach the values of 0.78 and -0.42}
\end{figure}

The density field $\rho(\v{x},t)$ is converted to a time fluctuating field
by subtracting the time average $\bar{\rho}(\v{x})$:
$\rho'(\v{x},t)=\rho(\v{x},t)-\bar{\rho}(\v{x})$.  In the LSE framework  an
approximation $\hat{\rho}(\v{x},t)=\alpha \bar{\phi}(t) + \beta
\bar{\psi}(t)$ to $\rho'(\v{x},t)$ is found that minimises the residual
error $\langle \rho' - \hat{\rho}\rangle_t$ \cite{Papoulis2002}, where
$\bar{\phi}(t)$, $\bar{\psi}(t)$ are the angles averaged over all the
frames at time $t$ and $\alpha$, $\beta$ are constants.
$\hat{\rho}(\v{x},t)$ represents the density time fluctuations correlated
with the angles.  It is calculated by solving for $\alpha$, $\beta$ the
system of linear equations, obtained from $\langle \rho'(\v{x},t)
\bar{\phi}(t)\rangle = \langle\{\alpha \bar{\phi}(t) + \beta
\bar{\psi}(t)\} \bar{\phi}(t)\rangle$ (and similarly for $\bar{\psi}(t)$)
in the assumption of the zero mean of the uncorrelated part of the density
fluctuations:
\begin{eqnarray*}
\alpha\langle \bar{\phi}(t)\bar{\phi}(t)\rangle + \beta \langle 
\bar{\psi}(t)\bar{\phi}(t)\rangle & = & \langle \rho'(\v{x},t) 
\bar{\phi}(t)\rangle \\
\alpha\langle \bar{\phi}(t)\bar{\psi}(t)\rangle + \beta \langle 
\bar{\psi}(t)\bar{\psi}(t)\rangle & = & \langle \rho'(\v{x},t) 
\bar{\psi}(t)\rangle.
\end{eqnarray*}


The density fluctuations field $\rho'(\v{x},t)$ and its part correlated 
with the peptide's conformation $\hat{\rho}(\v{x},t)$ are shown in 
Fig.~\ref{fig:dens_corr} for several representative time moments. 

The water fluctuations (right column) are significantly stronger just
before the transition ($2.4ps$) compared to the stable period ($\approx
100ps$). Very surprisingly the conditionally averaged water fluctuations
(left column) are virtually uncorrelated with the peptide at all times
except for the short period immediately before the transition (some
fluctuations are also noticeable as early as $\approx 20ps$ before the
transition, which agrees with the onset of the density decrease in
Fig.~\ref{fig:av_density}). Interestingly, at $0ps$, when the transition
process is complete, the water density becomes uncorrelated with the
peptide, similar to the stable periods.  However, the fluctuations of it
remain strong, only slightly weaker than at $2.4ps$.  We explain this
effect by the large inertia of the water shell, that needs relatively long
time for the fluctuations to settle down. The fact that these
post-transition water fluctuations are decoupled from the peptide
emphasizes the discovered phenomenon of strong water-peptide interactions
precisely during the transition process.

Summarising, we have found that (i) $\approx 5ps$ before the transition,
when the dihedral angles change the most, the water density significantly
reduces; (ii) the change of water density begins at $\approx 50ps$ before
the transition, 10 times earlier than the changes in the angles (iii)
during the transition the dynamics of water density becomes highly
correlated with the dynamics of the angles; and (iv) these correlations are
completely absent during the stable conformation periods.

We conclude that water and the peptide behave as an integral dynamical
system. During the conformational transition the peptide and the
surrounding water undergo transitions together.  This is in contrast to the
metastable periods when their dynamics is essentially decoupled.  The
transition is characterised by a more diffuse hydrogen bonds network of
water. The changes in the peptide are substantially delayed in time. Thus,
it is likely that water drives the whole process of conformational
transitions.

\end{document}